\documentclass[preprint, amsmath,amssymb,aps,prapplied,longbibliography]{revtex4-1}
\usepackage{graphicx}
\usepackage{dcolumn}
\usepackage{amsmath}
\usepackage{here}
\usepackage{bm}
\usepackage{color}

\begin{document}

\title{
  Low power switching of magnetization using enhanced magnetic
  anisotropy with short-voltage-pulse application
}

\author{R. Matsumoto}\email{rie-matsumoto@aist.go.jp}
\author{H. Imamura}\email{h-imamura@aist.go.jp}

\affiliation{ 
National Institute of Advanced Industrial Science and Technology (AIST),
Spintronics Research Center, Tsukuba, Ibaraki 305-8568, Japan
}

\date{\today}

\begin{abstract}
  A low power magnetization switching scheme based on the voltage control of
  magnetic anisotropy (VCMA) is proposed. In contrast to the
  conventional switching scheme using VCMA, where the
  magnetic anisotropy is eliminated during the voltage pulse, the
  magnetic anisotropy is enhanced to induce precession around the
  axis close to the easy axis. After turning off the voltage 
  proximately at a half of precession period, the magnetization relaxes to the opposite
  equilibrium direction. We perform numerical simulations and show
  that the pulse duration of the proposed switching scheme is
  as short as a few tens of pico seconds. Such a short pulse duration
  is beneficial for low power consumption 
  because of the reduction of
  energy loss by Joule heating.
\end{abstract}


\maketitle

\section{Introduction}
Low power consumption is a key requirement for modern computational devices.
Non-volatility is one of the core concepts to reduce 
power consumption for logics and memories
in normally-off computing
\cite{ando_fed_2001, ando_spin-transfer_2014, nakada_normally-off_2017}. 
Magnetoresistive random access memory (MRAM) is a promising
non-volatile memory that stores information 
associated with the direction of
magnetization in magnetic tunnel junctions (MTJs)
\cite{yuasa_giant_2004, parkin_giant_2004, djayaprawira_230_2005, yuasa_giant_2007,
kishi_lower-current_2008, kitagawa_impact_2012,
apalkov_magnetoresistive_2016, sbiaa_recent_2017, cai_high_2017}. 
In order to reduce
power consumption of MRAM, 
several types of writing schemes have been
developed. The currently used writing scheme is based on the
spin-transfer-torque (STT) switching phenomena, which were
proposed by
Slonczewski \cite{slonczewski_conductance_1989,slonczewski_current-driven_1996}
and independently by Berger \cite{berger_emission_1996}. The write
energy of STT-MRAM is of the order of 100 fJ/bit
\cite{kitagawa_impact_2012, cai_high_2017}, which is
still 2 orders of magnitude larger than that of static random-access
memory.

\begin{figure}[t]
\centerline{\includegraphics [width=\columnwidth] {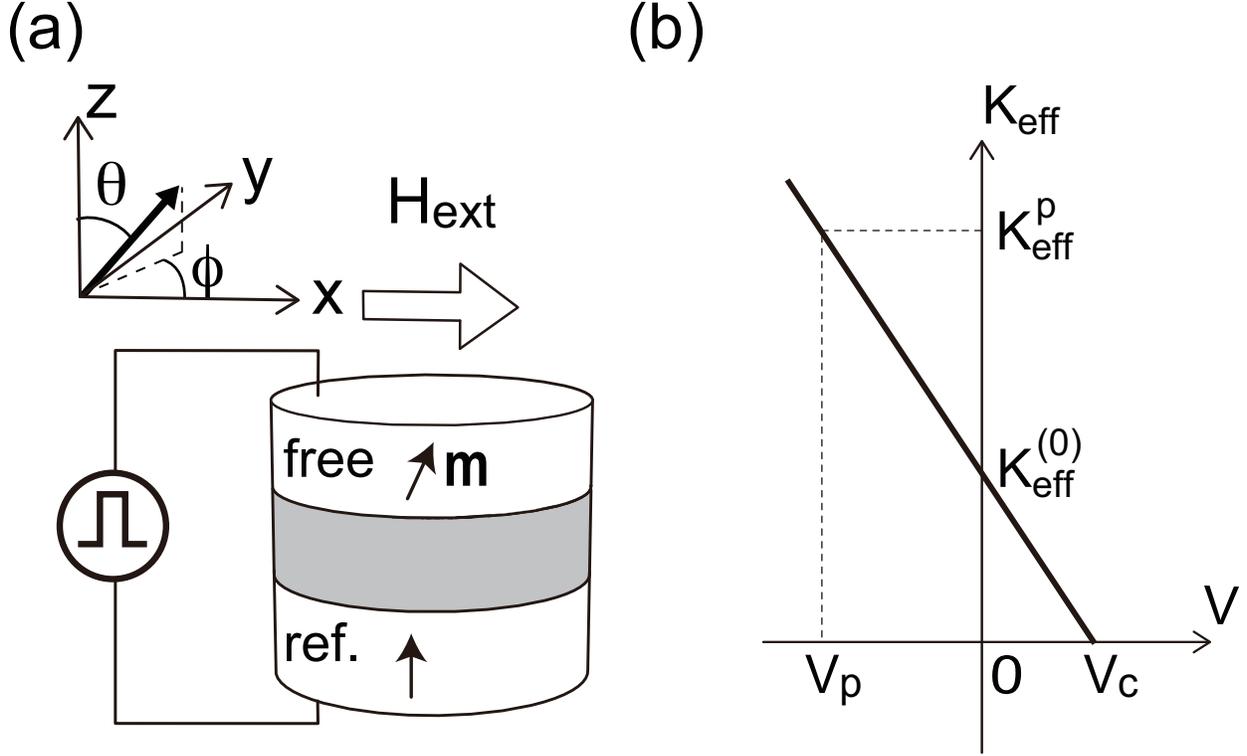}}
  \caption{
  \label{fig:schem} 
  (a)  Magnetic tunnel junction with circular cylinder shape,
  and definitions of Cartesian coordinates $(x, y, z)$, polar angle
  ($\theta$), and azimuthal angle ($\phi$).
  The $x-$axis is parallel to the direction of 
  the external in-plane magnetic field, ${\bm H}_{\rm ext}$.
  The unit vector ${\bm m} = (m_{x}$, $m_{y}$, $m_{z})$ 
  represents the direction of the magnetization in the free layer.
  The magnetization in the reference layer (ref.) is fixed to align in the
  positive $z-$direction.
  (b) The voltage ($V$) dependence of the effective perpendicular anisotropy constant ($K_{\rm eff}$).
  The effective anisotropy constant at $V=0$ is represented by
  $K_{\rm eff}^{(0)}$. It takes the value of
  $K_{\rm eff} = K_{\rm eff}^{\rm  p}$ at $V=V_{\rm p}$.
  }
\end{figure}

Discovery of the voltage control of magnetic anisotropy (VCMA) effect
\cite{weisheit_electric_2007, maruyama_large_2009, duan_surface_2008,
  nakamura_giant_2009, tsujikawa_finite_2009, nozaki_voltage-induced_2010,
  endo_electric-field_2010,
  nozaki_magnetization_2014, skowronski_perpendicular_2015, nozaki_large_2016,
  li_enhancement_2017}
paved the way for further reduction of write energy in MRAM.
The mechanism of VCMA in an MgO-based MTJ is considered to be the
combination of the selective electron or hole doping into the
$d$-electron orbitals and the induction of a magnetic dipole moment,
which affect the electron spin through spin-orbit interaction
\cite{duan_surface_2008,nakamura_giant_2009,tsujikawa_finite_2009,miwa_voltage_2017}.
The MRAM which uses the VCMA effect to switch magnetization is called
the voltage controlled MRAM (VC-MRAM)
\cite{shiota_induction_2012, shiota_pulse_2012,
kanai_electric_2012,shiota_evaluation_2016, grezes_ultra-low_2016, kanai_electric-field-induced_2016,
shiota_reduction_2017, 
matsumoto_voltage-induced_2018, yamamoto_thermally_2018, yamamoto_write-error_2019, 
matsumoto_voltage-induced_2019, imamura_impact_2019, matsumoto_methods_2019}.
The writing procedure of a conventional VC-MRAM is as follows.
The perpendicularly magnetized MTJ is subjected to an in-plane
external magnetic field ($H_{\rm ext}$) as shown in Fig. \ref{fig:schem}(a).
The magnetic anisotropy (MA) constant of the free layer can be controlled by
applying voltage ($V$) as shown in Fig. \ref{fig:schem}(b).
Here, $K_{\rm eff}$ is the effective perpendicular anisotropy constant
where the demagnetization energy is subtracted from the perpendicular anisotropy constant.
Throughout the paper, the superscript (0) indicates the quantities at $V = 0$.
The voltage pulse with critical amplitude $V_{\rm c}$
eliminates the MA and induces the precession of the magnetization around
the external magnetic field. By turning off the voltage at one half
period of precession, the magnetization switching completes.

The write energy of VC-MRAM is estimated from the Joule heating energy loss
during the pulse. Assuming that the voltage
pulse with amplitude $V$ and duration $t_{\rm p}$ is applied to the MTJ with
resistance $R$, the write energy is given by 
\begin{equation}
  E_{\rm J} = \frac{V^{2}}{R}t_{\rm p},
\end{equation}
To reduce the write energy, 
the VC-MRAM should be designed to have large
resistance and short pulse duration. The pulse duration is given by
a half period of precession as
\begin{align}
  \label{eq:period_Hext}
  t_{\rm p} = \frac{\pi (1+\alpha^{2}) }{\gamma H_{\rm ext}},
\end{align}
where $\alpha$ is the Gilbert damping constant and $\gamma$ is the
gyromagnetic ratio.  For example, $t_{\rm p}$ = 0.18 ns for $\alpha$ =
0.1 and $\mu_{0} H_{\rm ext}$ = 100 mT, where $\mu_{0}$ is the vacuum
permeability. Recently, Grezes $et$ $al$. demonstrated 
a very small write energy of 6 fJ/bit for the VC-MRAM with $R=330$ k$\Omega$ 
at $V$ = 1.96 V and $t_{\rm p}$ = 0.52 ns
\cite{grezes_ultra-low_2016}. The similar results were also obtained
independently by Kanai $et$
$al$. \cite{kanai_electric-field-induced_2016}.

It is difficult to use MTJ with huge $R$ to further reduce
write energy because the read time of the VC-MRAM increases 
with increase of $R$. 
Adopting a scheme of decreasing  
pulse duration
by increasing external magnetic field 
should also be avoided 
since the application of a
strong in-plane magnetic field $H_{\rm ext}$ deteriorates the thermal stability
factor defined as
\begin{align}
  \label{eq:delta}
  \Delta^{(0)}
  =
  \frac{
    \left(2 K_{\rm eff}^{(0)}
    - \mu_{0}M_{\rm s} H_{\rm ext}
    \right)^{2}
    V_{\rm F}
  }{4K_{\rm  eff}^{(0)} k_{\rm B}T},
\end{align}
where 
$k_{\rm B}$ is the Boltzmann constant, $T$ is the temperature,
$M_{\rm s}$ is the saturation magnetization, 
and $V_{\rm F}$ is the volume of the free layer.

In this paper, we propose another switching scheme which could reduce the pulse
duration and therefore the write energy of a VC-MRAM. The main
difference between the conventional scheme and the proposed switching scheme is
the polarity of the voltage pulse. Application of the voltage pulse
with the polarity that is opposite to the conventional switching can
enhance the magnetic anisotropy and induce the precession around the
axis close to the easy axis. After turning off the voltage 
proximately at a
half of precession period, the magnetization relaxes to the opposite
equilibrium direction and the switching completes.
We perform numerical simulations and demonstrate that the pulse
duration of the proposed switching scheme is as  short as a few tens
of pico seconds. We also evaluate the write error rate (WER) and
show that the WER is minimized if the pulse duration is about half the
period of precession similar to the conventional switching scheme.

\section{Theoretical model}
The system we consider is schematically shown in Fig. \ref{fig:schem}(a).
The macrospin model is employed to describe the magnetization dynamics.
The direction of the magnetization in the free layer is
represented by the unit vector ${\bm m} = (m_{x}$, $m_{y}$, $m_{z})
= (\sin \theta \cos \phi$, $\sin \theta \sin \phi$, $\cos \theta$), 
where $\theta$ and $\phi$ are the polar and azimuthal angles.
The $x$ axis is parallel to the direction of external in-plane
magnetic field ${\bm H}_{\rm ext}$.

The energy density of the free layer is given by
\begin{align}
  \label{eq:energy_density}
  {\cal E} (m_{x}, m_{y}, m_{z})
  = 
   -K_{\rm eff} m_{z}^{2} - \mu_{0} M_{\rm s} H_{\rm ext} m_{x},
\end{align}
The first term of Eq. \eqref{eq:energy_density} is the sum of the
shape, the bulk crystalline and the interfacial
anisotropies.  Owing to the VCMA effect, 
$K_{\rm eff}$ can be controlled
by application of $V$ as shown in Fig. \ref{fig:schem}(b).
Here $K_{\rm eff}^{(0)}$ represents the effective anisotropy constant
without the voltage application. We assume that $K_{\rm eff}$
decreases with increase of $V$ and vanishes at $V=V_{\rm c}$.
Applying the voltage $V_{\rm p} ( < 0)$ increases
$K_{\rm eff}$ to $K_{\rm eff}^{\rm p}$ and induces
the precessional motion of $\bm{m}$ around the effective magnetic
field. The effective field is given by 
$H_{\rm eff} = (H_{\rm ext},$ $0,$ $H_{\rm
  K} m_{z})$, where ${H}_{\rm K} =  2 K_{\rm eff}^{\rm p} / ( \mu_{0}
M_{\rm s} )$ is the anisotropy field.

The magnetization dynamics is simulated by solving 
the following Landau-Lifshitz-Gilbert equation \cite{brown_thermal_1963},
\begin{equation}
 \label{eq:LLG}
 \frac{{\rm d} {\bm m}}{{\rm d}t}
 = -\gamma_{0} {\bm m}\times
 \left(\bm{H}_{\rm eff} + \bm{h}\right)
 +\alpha
 \bm{m}\times
  \frac{{\rm d} {\bm m}}{{\rm d}t},
\end{equation}
where $\bm{h}$ represents the thermal agitation field satisfying the
following relations:
\begin{align}
  &\langle h_{\iota}(t)\rangle=0
  \\
  & \langle
h_{\iota}(t)h_{\kappa}(t') 
\rangle
= \frac{2\alpha k_{\rm B} T }{ \gamma_{0} \mu_{0} M_{\rm s} V_{\rm F} }\delta_{\iota\kappa}\delta(t-t').
\end{align}
Here  $\iota,\kappa=x,y,z$, 
and $\langle X \rangle$ denotes the statistical average of $X$.

Throughout this paper, we assume that the external field is $\mu_{0}
H_{\rm ext}$ = 100 mT and the saturation magnetization of the free
layer is $M_{\rm s}= 1400$ kA/m.
Also the radius of the junction area is assumed as $r=50$ nm and the
thickness of the free layer, $t_{\rm F}=1$ nm, and therefore the volume of
the free layer as $V_{\rm F}=\pi r^{2} t_{\rm F}=7854$ nm$^{3}$. 
The initial states are prepared by 10 ns relaxation from the
equilibrium direction at $T=0$, that is 
$(\theta^{(0)},$ $\phi^{(0)})  =  \left( \sin^{-1}  \left[ \mu_{0} M_{\rm s} H_{\rm ext}  / (2 K_{\rm eff}^{(0)} ) \right] \right.,$ 
$ 0 \Bigr)$
\cite{matsumoto_voltage-induced_2019}. 
The write error rates are calculated from
$10^{6}$ trials with 10 ns relaxation after the pulse.

\section{Results and discussions}
\begin{figure}[H]
  \centerline{
  \includegraphics [width=0.8\columnwidth] {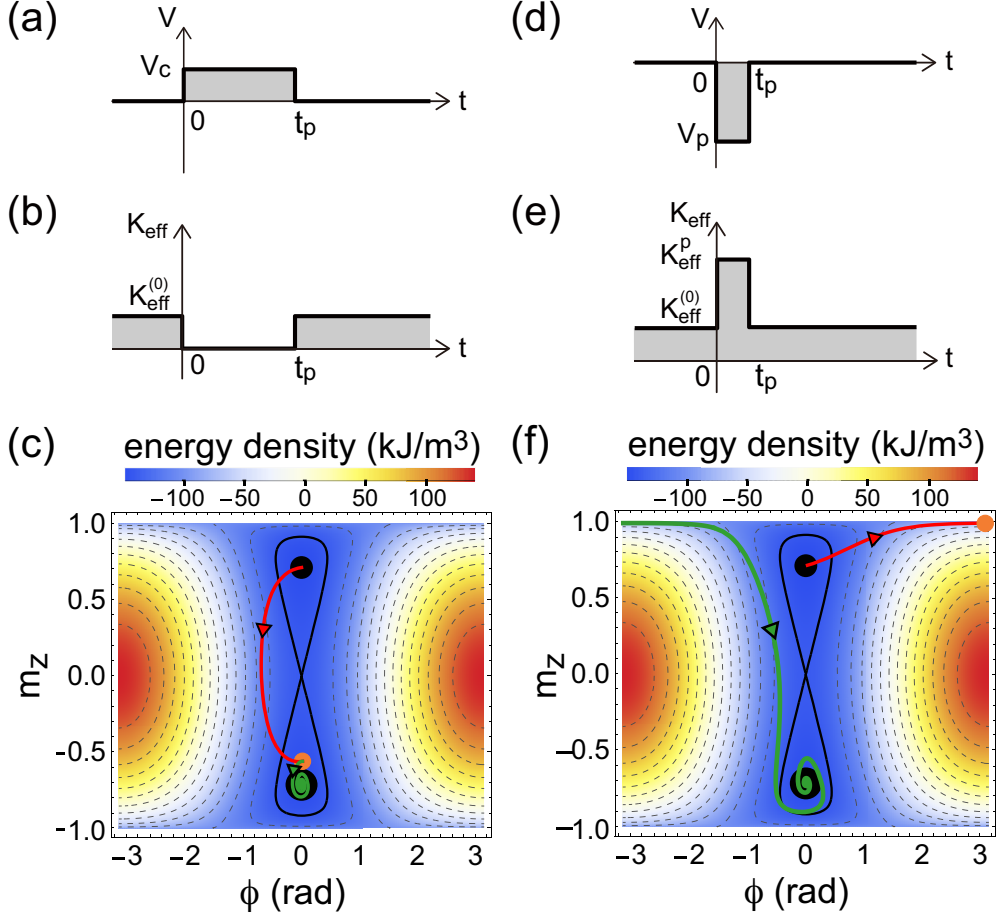}
  }
  \caption{
    \label{fig:t0} 
    (a) The shape of the voltage pulse for the {\em conventional switching
    scheme}. The amplitude including the polarity of the pulse and duration of the pulse are 
    $V_{\rm c}$ (positive value) and $t_{\rm p}$, respectively.
    (b) The corresponding time dependence of the effective anisotropy
    constant $K_{\rm eff}$. At $V=0$, it takes the value $K_{\rm eff}^{\rm (0)}$. 
    During the pulse, $K_{\rm eff} = 0$ because $V=V_{\rm c}$.
    (c) The color map of the energy density at $V=0$ on the
    $\phi-m_{z}$ plane. Thin black dotted curves represent energy
    contours. Thick black curves represent the energy contour crossing
    $\bm{m}=(1,$ $0,$ $0)$.
    The trajectory of $\bm{m}$ during and after the pulse
    are shown by the red and green curves, respectively. The direction
    of the trajectory is indicated by the triangle. The orange circle
    represents the direction of  ${\bm m}$ at the end of the pulse.
    We assume that $\alpha$ = 0.1.
    (d) The shape of the voltage pulse for the {\em proposed switching scheme}.
    The polarity is negative, i. e. $V_{\rm p}<0$, to enhance $K_{\rm eff}$.
    (e) The corresponding time dependence of the effective anisotropy
    constant.  During the pulse, it is enhanced to $K_{\rm eff}^{\rm p}$.
    (f) The color map of the energy density at $V=0$ on the
    $\phi-m_{z}$ plane.  We assume that $K_{\rm eff}^{\rm p}$ = 400
     kJ/m$^{3}$ and $\alpha$ = 0.21. The symbols are the same as those
     in Panel (c).
     Please note that the left and right boundaries at  $\phi =
    \pm \pi$ represent the same direction of $\bm{m}$.
}
\end{figure}

First we show the difference between the mechanisms of the conventional
voltage controlled switching and the proposed switching that utilizes
the enhancement of the magnetic anisotropy.
This can be accomplished by analyzing the switching
trajectories at $T=0$.
Figures \ref{fig:t0}(a) and (b) show the shape of the voltage
pulse and the corresponding time dependence of the effective
anisotropy constant for the conventional voltage controlled switching.
The induced switching dynamics of $\bm{m}$ at $T=0$ is shown in
Figs. \ref{fig:t0}(c) together with the color map of the energy
density of Eq. \eqref{eq:energy_density} at $V=0$.
Thin black dotted curves represent energy contours. Thick black curves
represent the energy contour crossing
$\bm{m}=(1,$ $0,$ $0)$. The initial direction of the magnetization is the
equilibrium direction with $m_{z}>0$ indicated by the black circle,
which we call as the ``up state''.  

In Figs. \ref{fig:t0}(a), (b) and (c),
application of the voltage pulse with $V_{\rm c}$
eliminates the magnetic anisotropy and induce the precession of $\bm{m}$
around the external magnetic field as represented by the red
curve.  After turning off the voltage at one-half period of precession, 
the magnetization starts to relax from the point indicated by the orange
circle
to the other equilibrium
direction with $m_{z}<0$, i. e. the ``down state'', indicated by the
black circle. 
Note that the black circle at $m_{z}<0$ is illustrated 
under the green curve. 
The switching is thus completed as represented by the green
curve.

Figures \ref{fig:t0}(d) and (e) show the shape of the voltage pulse
and the corresponding time dependence of the effective anisotropy
constant for the switching using the enhanced $K_{\rm eff}$.
The induced switching dynamics of $\bm{m}$ at $T=0$ is shown in
Fig. \ref{fig:t0}(f) together with the color map of the energy
density at $V$ = 0. 
The initial state is the up state indicated by the
black circle at $m_{z}>0$. Application of the voltage pulse with $V_{\rm p} (< 0)$ 
enhances the effective anisotropy constant from $K_{\rm eff}^{(0)}$ to
$K_{\rm eff}^{\rm p}$ and induce the precession of $\bm{m}$
around the effective magnetic field as represented by the red curve.
The value of $K_{\rm eff}^{\rm p}$ is assumed to be 400
kJ/m$^{3}$, which gives the anisotropy field of $\mu_{0} H_{\rm K}$ = 570
mT. The effective field is nearly parallel to the easy axis or the $z$
axis because the directional cosine of the effective field relative to the
easy axis is 0.98. 
The voltage is turned off at about a half period
of the precession, and the magnetization reaches the point, 
$\phi \simeq \pi$ indicated by the orange circle. As will be shown later, 
the write error rate (WER) is minimized if the pulse duration is set about
half the period of precession. After turning off the pulse, the
magnetization relaxes to the down state and completes the switching as
shown by the green curve. The proposed switching scheme does not
reduce the thermal stability factor of Eq. \eqref{eq:delta} 
because it
just enhances $K_{\rm eff}$ during the voltage pulse.

\begin{figure}[H]
\includegraphics [width=\columnwidth] {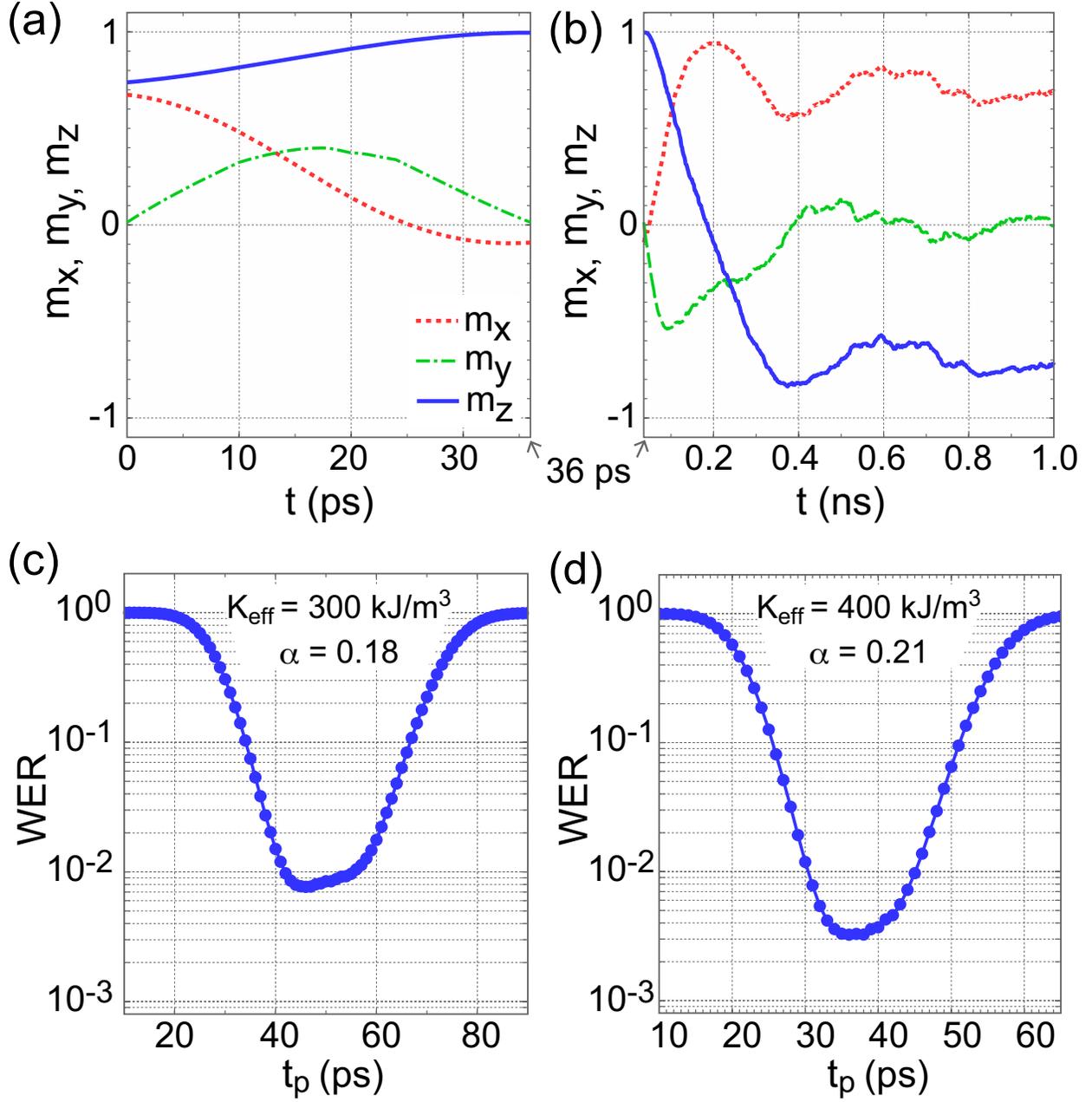}
\caption{\label{fig:mm} 
  (a) The Cartesian components of ${\bm m} = (m_{x}$, $m_{y}$,
  $m_{z})$ of a typical switching trajectory 
  are plotted as a function of time
  during the
  pulse at $T$ = 300 K.
  $K_{\rm eff}^{\rm p}$ = 400 kJ /m$^{3}$ and $\alpha$ = 0.21.
  The unit of the horizontal axis is ps.  
  (b) The same as (a) after the pulse. The unit of the horizontal axis is ns.  
  (c) The pulse duration dependence of the WER at $T$ = 300 K for
  $K_{\rm eff}^{\rm p}$ = 300 kJ /m$^{3}$ and $\alpha$ = 0.18.
  (d) The same as (c) for $K_{\rm eff}^{\rm p}$ = 400 kJ /m$^{3}$ and
  $\alpha$ = 0.21.
  }
\end{figure}

Next we discuss the switching properties of the proposed switching
scheme at $T$ = 300 K by analyzing the numerical simulations results.
The time evolution of the Cartesian components of ${\bm m}$ for a
typical switching trajectory during the pulse are
shown in Fig. \ref{fig:mm}(a). The value of $K_{\rm eff}^{(0)} = 100$ kJ/m$^{3}$
and $\alpha$ are 
the same as in Fig. \ref{fig:t0}(f),
$K_{\rm eff}^{\rm p}$ = 400 kJ/m$^{3}$ and $\alpha$ = 0.21.
During the pulse duration, 
$m_{z}$ increases with the
increase of time because the effective anisotropy constant is
enhanced. The shapes of $m_{x}$ and $m_{y}$ are very similar to
the cosine and sine functions, respectively, 
because 
$\bm{m}$ precesses around the effective field which is almost parallel
to the $z$ axis. 
Figure \ref{fig:mm}(b) shows the time evolution of $m_{x}$, $m_{y}$
and $m_{z}$ after the pulse. Please note that the horizontal axis is
in unit of ns. $m_{z}$ monotonically decreases with the increase of time
and the switching completes at around 0.4 ns.

Figures \ref{fig:mm}(c) and (d) show the pulse duration, $t_{\rm p}$,
dependence of the write error rate (WER) for different
values of $K_{\rm eff}^{\rm p}$ and $\alpha$.
The parameters are  $K_{\rm eff}^{\rm p}$ = 300 kJ /m$^{3}$ and
$\alpha$ = 0.18 for Fig. \ref{fig:mm}(c), 
and $K_{\rm eff}^{\rm p}$ =
400 kJ /m$^{3}$ and $\alpha$ = 0.21 for Fig. \ref{fig:mm}(d).
In Fig. \ref{fig:mm}(c), the WER takes a minimum value of $7.6 \times
10^{-3}$  at $t_{\rm p}$ = 46 ps.
In Fig. \ref{fig:mm}(d), the WER takes a minimum value of $3.2 \times
10^{-3}$  at $t_{\rm p}$ = 36 ps. These optimal values of $t_{\rm p}$ at which
the WER is minimized are almost the same as one half period of precession around
$H_{\rm eff}$.

\section{conclusion}
In summary, we propose a low power switching scheme of magnetization using
enhanced magnetic anisotropy by applying a short voltage pulse.
The proposed switching scheme can reduce the pulse duration and
therefore the write energy substantially without  deteriorating 
thermal stability.
We perform numerical simulations and show that the pulse duration of
the proposed switching scheme is as short as a few
tens of pico seconds.  We also calculated the pulse duration
dependence of the WER, 
and showed that the optimal values of $t_{\rm
  p}$ at which the WER is minimized are nearly half the period of
precession around the effective field.

\acknowledgements
This work was partly supported by JSPS KAKENHI Grant No. JP19K05259
and 19H01108.


%

\end{document}